\documentclass[12pt]{article}

\usepackage{scicite}

\usepackage{times}

\usepackage{graphicx}
\graphicspath{ {./images/}}

\newenvironment{sciabstract}{%
\begin{quote} \bf}
{\end{quote}}

\newcounter{lastnote}

\title{Circumventing size-bandwidth limits in imaging with flat lenses}

\author
{Apratim Majumder,$^{1}$ Monjurul Meem,$^{1}$ Nicole Brimhall,$^{2}$ Rajesh Menon,$^{1,2\ast}$\\
\\
\normalsize{$^{1}$Department of Electrical and Computer Engineering, University of Utah,}\\
\normalsize{50 Central Campus Dr, Salt Lake City, UT 84112, USA}\\
\normalsize{$^{2}$Oblate Optics, Inc., San Diego, CA 92130, USA}\\
\\
\normalsize{$^\ast$To whom correspondence should be addressed; E-mail:  rmenon@eng.utah.edu.}
}

\date{}

\begin{document} 


\baselineskip24pt


\maketitle


\begin{sciabstract}
  Recent theoretical work suggested upper bounds on the operating bandwidths of flat lenses. Here, we show how these bounds can be circumvented via a multi-level diffractive lens (MDL) of diameter = 100\thinspace mm, focal length = 200\thinspace mm, device thickness = 2.4\boldmath$\mu$\thinspace m and operating bandwidth from \boldmath$\lambda$ = 400\thinspace nm to 800\thinspace nm. We further combine the MDL with a refractive lens to demonstrate a hybrid telescope. By appealing to coherence theory, we show that the upper bound on relative bandwidth is surprisingly independent of lens diameter or numerical aperture, but is only limited by the bandwidth of the image sensor. Since large-area achromatic flat lenses produce significant reductions in weight over their refractive counterparts, these calculations and experiments open up opportunities for very large scale diffractive and diffractive-refractive telescopes.
\end{sciabstract}



Lightweight, large-area optics could be extremely useful for airborne and space-based imaging and sensing systems ({\it 1, 2\/}). Although there has been tremendous progress in the field of ultra-lightweight flat lenses, computational and fabrication constraints have generally prevented demonstrations of large apertures. Previous efforts included the design and preliminary demonstrations of diffractive flat lenses, but these suffered from narrow operating bandwidths ({\it 3\/}). Furthermore, full aperture experimental demonstrations were not completed. Due to their low weight, diffractive flat lenses hold great promise for large-aperture space telescopes with potential applications in imaging exoplanets ({\it 4, 5\/}). Recent work has suggested the use of holographic elements for large-aperture spectral imaging ({\it 6\/}). Advances in subwavelength diffractive optics such as metalenses have resulted in 20\thinspace mm diameter flat lenses, but again only for a narrow bandwidth ({\it 4\/}).

Recent theoretical work has noted a fundamental limit on the size-bandwidth product of a flat lens  as function of numerical aperture (NA) ({\it 4\/}). Intuitively, this limit arises from the time delay experienced by the photons traversing the edges of the lens when they reach the focal point compared to that experienced by those photons traversing the optical axis (see inset Fig. 1a). It was pointed out that this time delay imposes a fundamental limit on the bandwidth of the illumination that can be focused by the lens. This time delay is given by $\Delta T=\frac{\sqrt{R^2+f^2}-f}{c}$, where $R$ and $f$ are the radius and focal length of the lens, and $c$ is the speed of light. In order to ensure coherent interactions between these photons, this time delay then imparts an upper bound on the maximum bandwidth of the illumination as:  $\Delta T\times\Delta\omega\le\kappa$, where $\kappa$ is a device-dependent constant, typically $\sim$\thinspace 10. Assuming R = 50\thinspace mm, f = 200\thinspace mm, we get $\Delta T = 0.21$\thinspace ns. And the upper bound on bandwidth becomes (for $\kappa \sim$\thinspace 10)   $\Delta\omega\ <\ 48\times{10}^9$\thinspace rad/s. 

In this work, we exploit a loophole specific to imaging to overcome this limit, where we first recognize that coherent interactions between photons transiting the edges of the lens and those along the optical axis are not required. As a result, the bandwidth of an imaging system is not limited by the bandwidth of the lens, but by the much smaller bandwidth of the sensor (and associated electronics) used to record the image. Specifically, the integration time of the image sensor is the key limiting factor, since the goal in imaging is to record the photons transiting the edges and the center of the lens within the same frame.  To illustrate this limit, we plot the time delay between photons arriving from the edges of the lens compared to the ones arriving along the optical axis for different lens diameters and numerical apertures (NAs) in Fig. 1a. As expected, this time delay increases with NA and with diameter. However, even for extremely fast detectors, say 1\thinspace MHz, lens diameter of 1000\thinspace m with NA as high as 0.85 is possible. For ultrafast detectors (1\thinspace GHz), the lens diameter is limited to approximately 1\thinspace m at NA $\sim$\thinspace 0.9. Note that these limits are independent of the bandwidth of the lens itself. The pixels of the sensor accumulate the photons during the exposure time period. Imaging is possible as long as the photons passing the edges of the lens arrive within the same interval of integration as those along the optical axis. Here, we exploit this concept to demonstrate a visible-band flat lens (Figs. 1b and 1c) with diameter = 100\thinspace mm and focal length = 200\thinspace mm, which corresponds to NA = 0.24 (f/2). The focal length of such a lens remains fixed (200\thinspace mm) over an operating bandwidth of 400\thinspace nm to 800\thinspace nm, which corresponds to a radial-frequency bandwidth, $\Delta\omega\ =\ 2.4\times{10}^{15}$\thinspace rad/s, many orders of magnitude larger than the strict limit dictated above. We note that bandwidth of this lens is far below the limit illustrated in Fig. 1a, suggesting significant room for improvement from our current demonstration.

\renewcommand{\figurename}{\bf Fig.}
\begin{figure}[htbp]
\includegraphics[width=\textwidth]{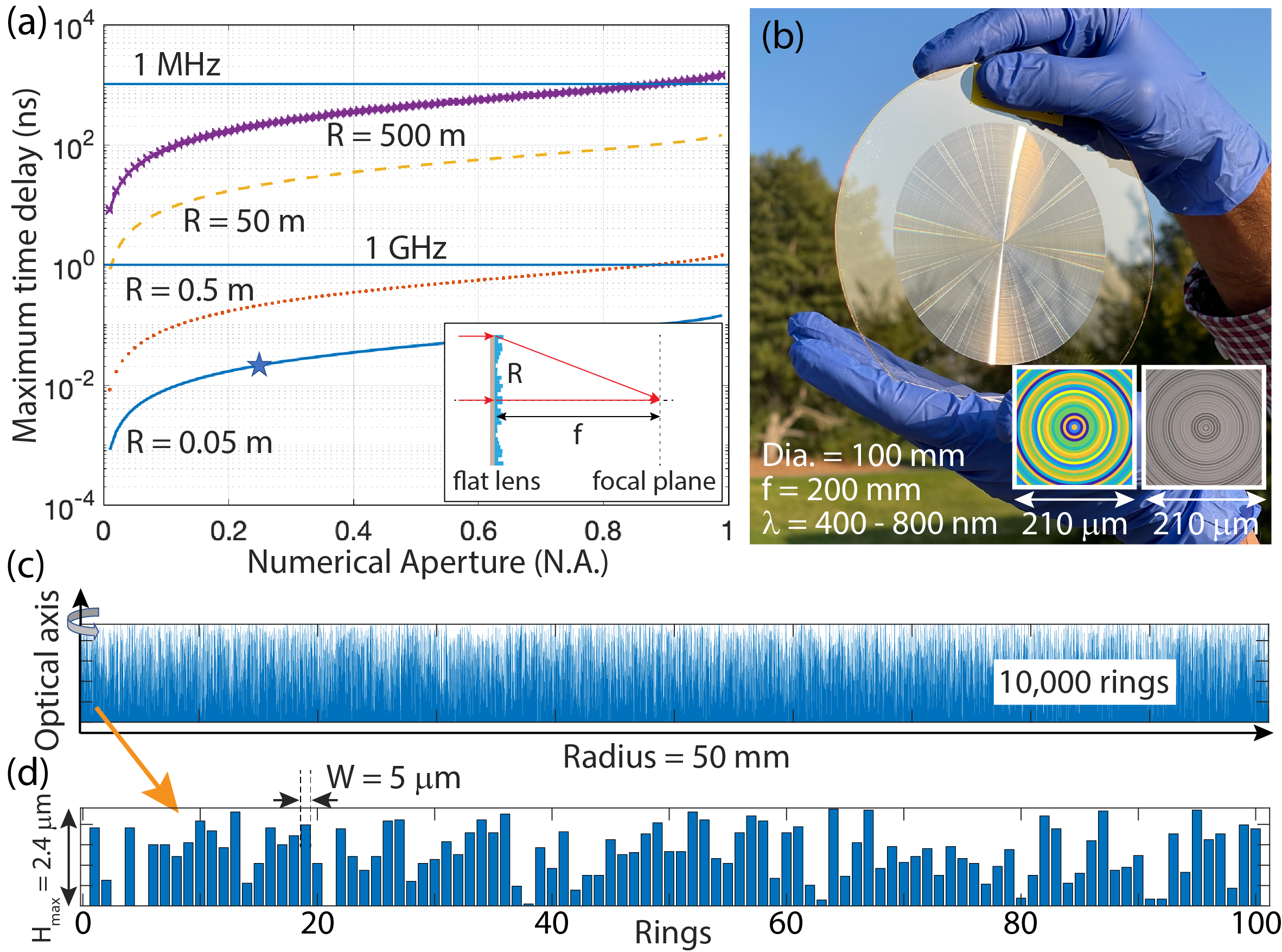}
\centering
\caption{ \normalfont Size-bandwidth limits of flat lenses. (a) Time delay (ns) experienced by marginal rays relative to on-axis rays when a flat lens is illuminated by a collimated plane wave at normal incidence as function of numerical aperture (NA). Inset shows the relevant geometry. Horizontal lines labelled 1\thinspace MHz and 1\thinspace GHz denote the limits on NA dictated by the bandwidth of the image sensor for different lens radii. (b) Photograph of the MDL with diameter = 100\thinspace mm and focal length = 200\thinspace mm (NA\thinspace =\thinspace  0.24). Bottom insets show the design (left) and optical micrograph (right) of the central rings of the MDL. (c) Design of a multilevel diffractive lens (MDL) with diameter = 100\thinspace mm and focal length of 200\thinspace mm. (d) Bottom inset shows magnified view of the central 100 rings.}
\end{figure}

Increasing the bandwidth of diffractive lenses have been subject of research for many years using approaches including diffractive-refractive hybrid systems ({\it 9\/}). multi-order lenses ({\it 10\/}), transformation-optics lenses ({\it 11\/}), and more recently, multi-level diffractive lenses ({\it 12\/}). Recent design of a 240\thinspace mm-diameter harmonic diffractive lens for the astronomical R band (589-727nm) has revived interest in these approaches for weight reduction in space-based telescopes ({\it 13\/}). Although dispersion engineering via subwavelength features is promising for broadband flat lenses, the fabrication complexity and the requirement of high-refractive index materials has limited demonstrations to very small diameters ({\it 14 -- 16\/}). 

\paragraph{Design} The MDL is comprised of concentric rings of fixed width and varying heights. We exploit the cylindrical symmetry of this geometry to implement the scalar-diffraction model using the Hankel transform, which dramatically reduces computational complexity ({\it 17\/}). A gradient-assisted direct-binary-search algorithm is used to vary the heights of the constituent rings with the goal of minimizing the difference between the simulated point-spread function (PSF) and the diffraction-limited Airy function for each wavelength sample, averaged over the bandwidth of interest  ({\it 18\/}). In order to simplify fabrication, we limited the ring width to 5\thinspace$\mu$m. This resulted in a total of 10,000 rings. A second MDL with ring width of 2.5\thinspace$\mu$m (20,000 rings) was also fabricated with slightly smaller bandwidth. Although the designed bandwidth of this second lens was 450\thinspace nm to 650\thinspace nm, we noted that the lens was achromatic from 450\thinspace nm to 800\thinspace nm (see section 1 of supplement). In both cases, the ring heights varied between 0 and 2.4\thinspace$\mu$m, which was dictated by our grayscale lithography process. The measured dispersion curve of a positive-tone photoresist (Shipley S1813, Microchem) was also used in the design. We further note that iterative optimization-based inverse design is now commonly used in all flat-lens implementations including metalenses ({\it 19\/}).

\paragraph{Fabrication} The MDL geometry was patterned in the photoresist layer (maximum thickness = 2.4\thinspace$\mu$m) on top of a double-side polished sodalime glass wafer (diameter = 150\thinspace mm, thickness = 550\thinspace$\mu$m). Patterning was performed using optical grayscale lithography (see section 2 of supplement). The fabricated device is shown in Fig. 1c. An optical micrograph of the central rings is compared to the design in the inset panels. A photograph of the MDL along with a plano-convex refractive singlet (diameter = 100\thinspace mm, focal length = 200\thinspace mm at $\lambda = 587.6$\thinspace nm, Edmund Optics, Model 27-501) is shown in Fig. S3.1 (supplement). The total weight of the MDL is 25\thinspace g, which is dominated by the weight of the glass wafer. Note that the wafer has no optical function, merely serving as a mechanical support. In comparison, a 100\thinspace mm-diameter commercial plano-convex refractive singlet weighs $\sim 211$\thinspace g. But we emphasize that the refractive singlet is, in fact not achromatic. A fully achromatic refractive imaging system will require multiple aspheric lenses and will be considerably heavier, and more expensive. Careful measurements of the height distribution of the fabricated rings confirm that the standard deviation of the height errors (compared to the designed values) is $\sim120$\thinspace nm (see section 2.1 in the supplement).

\paragraph {Experiments} We first measured the point-spread functions (PSFs) of the lenses as function of incident wavelength. Details of the experiment are provided in section 4 the supplement. Briefly, the setup consisted of a wavelength-tunable collimated beam (supercontinuum source SuperK FIU-15, NKT Photonics, connected to a SuperK VARIA filter, NKT Photonics), which was expanded by a series of negative lenses over $\sim 1.5$\thinspace m to produce a large diameter (approx. 230\thinspace mm) beam. The small divergence in the resulting beam caused a slight shift in the focal length of the MDL from 200\thinspace mm to $\sim 230$\thinspace mm. This was confirmed using a Shack-Hartmann sensor by measuring the focal length of the commercial plano-convex lens. The PSFs were recorded directly on a monochrome CMOS sensor (The Imaging Source, DMM 27UP031-ML, pixel pitch = 2.2\thinspace$\mu$m) and the results are summarized in Fig. 2a.  The distance between the lens and the sensor was fixed at 230\thinspace mm for all measurements. The MDL clearly produces focused spots at all wavelengths from 400\thinspace nm to 800\thinspace nm. Corresponding measurements for the refractive lens are included in the supplement confirming that the refractive lens is not achromatic as expected (see section 3 in the supplement). Fig. 2b shows the full-width-at-half-maximum (FWHM) of the focal spot produced by the MDL as shown in Fig. 2a, with respect to wavelength. Fig. 2c shows a plot of the Strehl ratio versus wavelength (average value = 0.38). The corresponding plots for the refractive lens are in section 3 (supplement). We note that the average Strehl ratio of the MDL is higher than that of the refractive singlet (at its design wavelength). The FHWM of the MDL PSFs are similar to those of the refractive lens. The background of the MDL reaches the average dark-frame value within a radius of $\sim 1.5$\thinspace mm from the peak. In comparison, the refractive singlet is clearly not achromatic. In section 8 of the supplement, we analyzed the PSFs of a commercial achromatic reractive lens (Thorlabs Model AC127-025-A). Both simulations (Fig. S8.1) and experiments (Fig. S8.2) confirmed that the MDL outperforms the achromat. As discussed later. we hypothesize that the differences in the background for the simulated and measured PSFs of the MDL are due to the 15\thinspace nm bandwidth of the experiment (smallest achievable value in our measurement setup). The off-axis PSFs of the MDL were also characterized, and we confirmed a field of view of approximately 4$^{\circ}$ (see Fig. 3a and details in section 4 of supplement).

\begin{figure}[htbp]
\includegraphics[width=\textwidth]{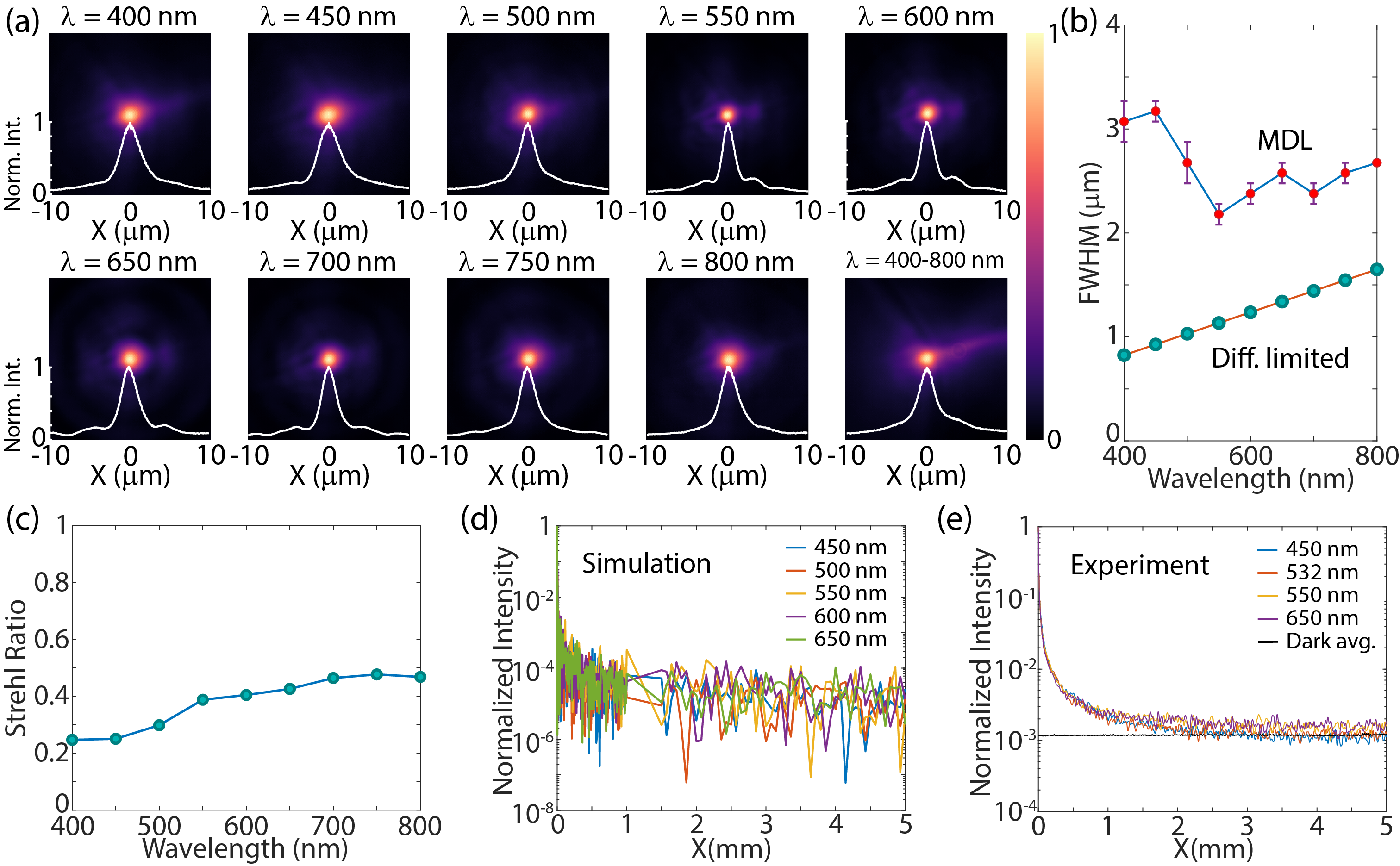}
\centering
\caption{ \normalfont Characterization of focusing. (a) Measured point-spread functions (PSFs) for various central wavelengths (bandwidth = 15\thinspace nm) and broadband illumination (400 - 800\thinspace nm). (b) Plot showing full-width-at-half-maximum (FWHM) of the measured PSFs in (a) with respect to wavelength as well as the corresponding diffraction limited spot size. (c) Strehl Ratio as function of wavelength. (d) Simulated and (e) measured radial cross-sections of a subset of the PSFs to emphasize the background level. The measured sensor dark level is shown as black line. The distance between the MDL and the sensor is fixed at 230\thinspace mm for all experiments. This is slightly longer than the designed focal length of 200\thinspace mm due to the small curvature of the incident beam.}
\end{figure}

Images of a USAF resolution chart were recorded using the MDL with the same sensor (Figs. 3b and 3c). The object and image distances were 400\thinspace mm and 400\thinspace mm in Fig. 3b, and 216\thinspace mm and 2735\thinspace mm in Fig. 3c, respectively. The smallest resolved lines are of Group 7, Element 4 corresponding to 118\thinspace line-pairs/mm. All images were processed using standard deblurring techniques using ImageJ (see section 5 in supplement). In each case, the exposure time was adjusted to ensure that the frames were not saturated.

Next, we took the camera (MDL + image sensor) outside to photograph the moon (details in section 5 of the supplement). The captured image of the moon (November 30, 2020 in Salt Lake City, UT) in shown in bottom panel of Fig. 3d.

\begin{figure}[htbp]
\includegraphics[width=\textwidth]{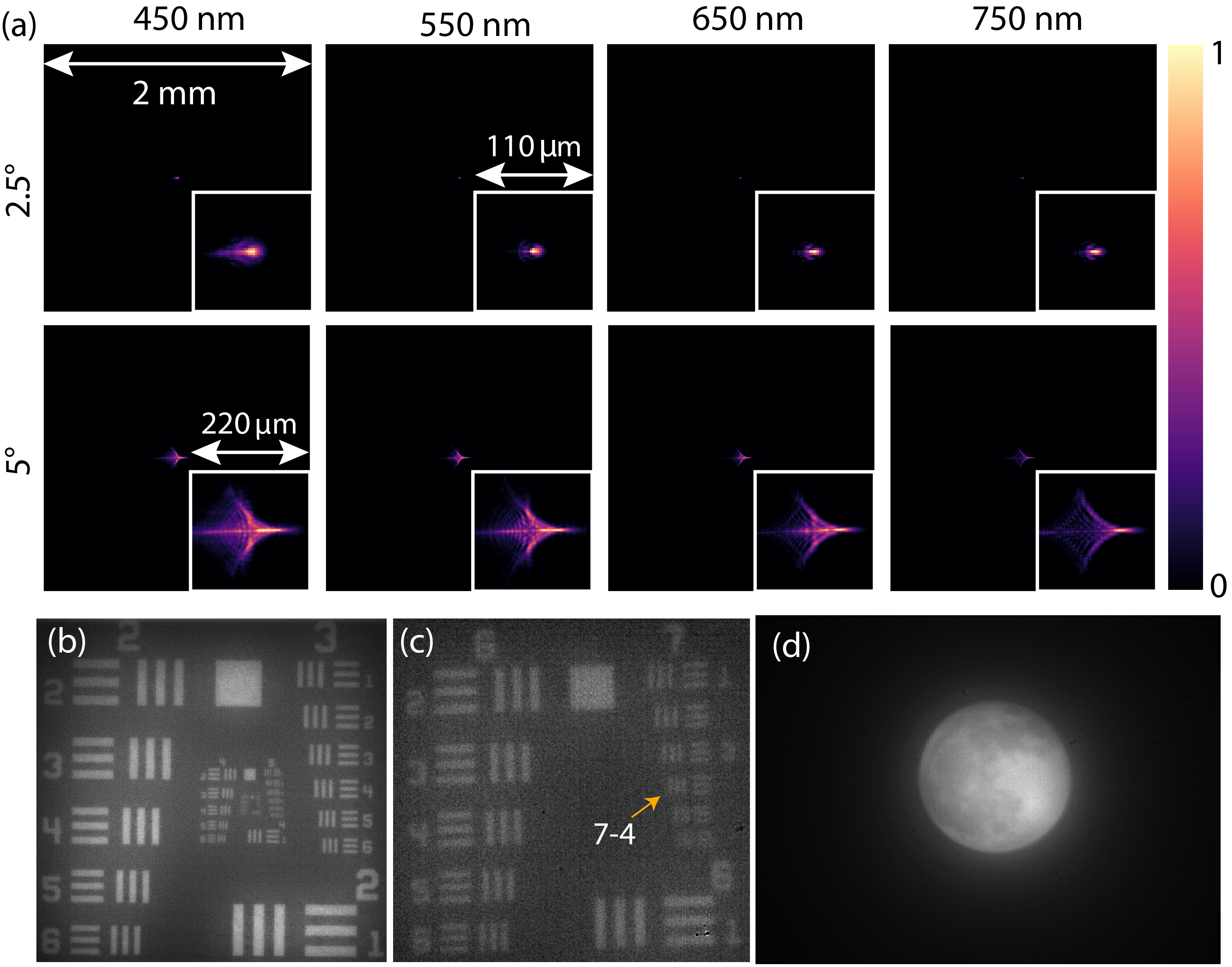}
\centering
\caption{ \normalfont Imaging performance of the MDL. (a) Recorded off-axis PSFs under various narrow-band illuminations (450\thinspace nm to 750\thinspace nm, bandwidth = 15\thinspace nm). (b) Image of USAF resolution chart with object distance = 400\thinspace mm and image distance = 400\thinspace mm. (c) Magnified image of the resolution chart with object distance = 216\thinspace mm and image distance = 2735\thinspace mm. The smallest resolved lines correspond to 181\thinspace lp/mm (Group 7, Element 4). (d) Photograph of the moon captured with MDL-sensor distance = 200\thinspace mm. }
\end{figure}

Finally, we built a hybrid Keplerian telescope comprised of the MDL as the primary lens and an achromatic refractive lens (diameter = 12.5\thinspace mm, focal length = 25\thinspace mm, Model Thorlabs AC127-025A) as the secondary element (Fig. 4a). The distance between the two lenses was 235\thinspace mm, while that from the secondary lens to the sensor was 87\thinspace mm (see details in section 6 of supplement). The recorded images of the resolution chart and that of the moon are in Figs. 4b and 4c, respectively. These images clearly show the potential of the MDL to replace components within conventional imaging systems to reduce overall weight. 

\begin{figure}[htbp]
\includegraphics[width=\textwidth]{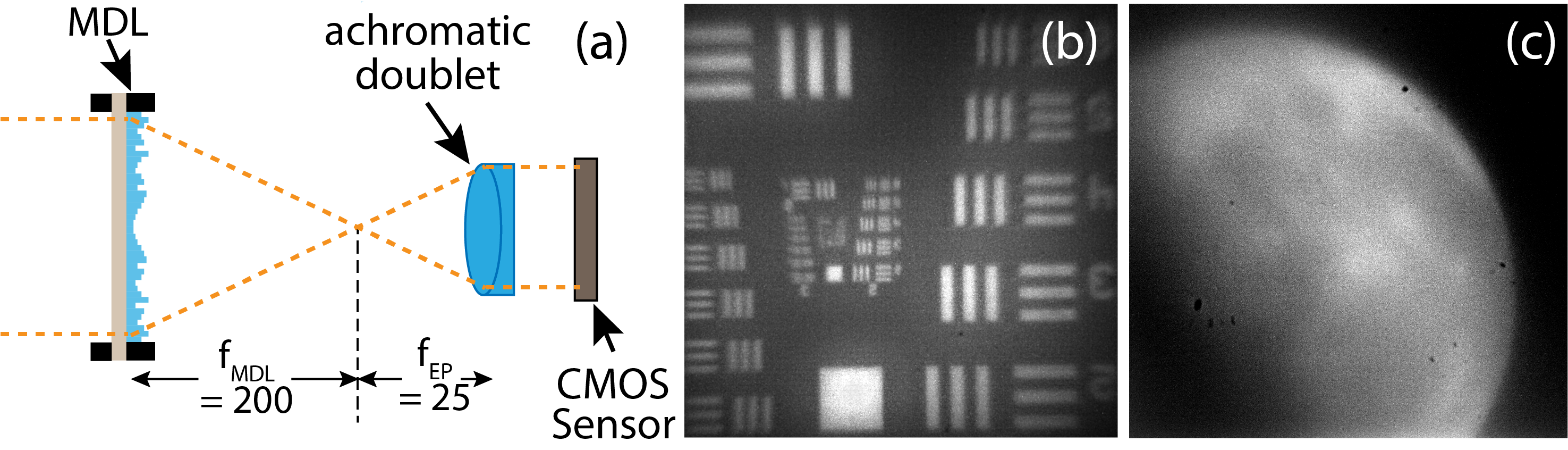}
\centering
\caption{ \normalfont Hybrid diffractive-refractive telescope. (a) Combining MDL with a refractive lens to create a hybrid Keplerian telescope. Magnified images captured of (b) a resolution chart and (c) the Moon. }
\end{figure}

\paragraph{Discussion} The MDL design was performed with discrete wavelength spacing of 25\thinspace nm from 400\thinspace nm to 800\thinspace nm. At these sampled wavelengths, the focusing performance is good (see Fig. 2 and Fig. S7.1). However, we observed that for wavelengths that were not sampled during design, the point-spread functions exhibit higher background (section 7 of supplement). Since the minimum bandwidth of our illumination system is 15\thinspace nm, this degrades the overall PSF of, even the narrowband measurements. We confirmed this by simulating the PSF for a continuous 15\thinspace nm spectrum with 1\thinspace nm spectral samples, and showed that it agrees well with the experimental data (Fig. S7.10). This accounts for somewhat increased background in the measured PSFs. It is also noted that such spectral dependencies could potentially be used for spectral imaging ({\it 6, 20\/}). The ideal ring-width should be less than the diffraction-limited FWHM of the smallest wavelength, which would be 0.5 x 400\thinspace nm / 0.24 = 833\thinspace nm. Since we used a much larger ring-width (5\thinspace$\mu$m or 2.5\thinspace$\mu$m), higher diffraction orders are required for focusing. This gives rise to not only increased background (and consequent reduction in Strehl ratio), but also an extended depth of focus (see Figs. S3.4 and section 7 in supplement). We speculate that by using ring-widths smaller than the diffraction limit as well with over-sampling in spectrum, it may be possible to overcome this spectral dependence, and elucidating these limits will be subject of future research.

One can derive a simpler, but weaker bound on the lens bandwidth by appealing to the fact that for high-contrast interference, the smallest optical-path-length difference must be larger than the coherence length of the illumination ({\it 21\/}). The coherence length, $l_c$ is approximately related to the bandwidth, $\Delta\lambda$ and the center wavelength, $\lambda_0$ as  $l_c\sim\frac{{\lambda_0}^2}{\Delta\lambda}$. The smallest path-length difference between two points (spaced by the ring-width, $w$, on the MDL with radius, $R$ and focal length, $f$) and the focus assuming plane-wave illumination is given by: $\sqrt{\left(R+w\right)^2+f^2}-\sqrt{R^2+f^2}$. In order to ensure interference between light traversing these two points and the focus, the coherence length of illumination must be greater than this path-length difference, {\it i.e.},  $l_c > \sqrt{\left(R+w\right)^2+f^2} - \sqrt{R^2+f^2}$. Putting these together, we can derive a simple upper bound on relative bandwidth: $\frac{\Delta\lambda}{\lambda_0}\le \frac{\lambda_0}{\sqrt{\left(R+w\right)^2+f^2}-\sqrt{R^2+f^2}}$. Since $w \ll\left\{R,f\right\}$, we can simplify the bound as $\frac{\Delta\lambda}{\lambda_0}\le \frac{\lambda_0\sqrt{R^2+f^2}}{Rw}$. If we further note that $w > \frac{min\left\{\lambda\right\}}{2NA} \ge \frac{\lambda_0}{2NA}$ to ensure diffraction into non-zero orders, we obtain the bound as $\frac{\Delta\lambda}{\lambda_0}\le 2$, which is independent of NA or the diameter of the lens. Since this is the same as the theoretical upper limit of relative bandwidth, $\frac{\Delta\lambda}{\lambda_0} = 2\times\frac{\lambda_{max}-\lambda_{min}}{\lambda_{max}+\lambda_{min}} \le 2$, we conclude that coherence imparts no fundamental limit on the operating bandwidth of a lens. Rather, it is the recording device (image sensor) that imparts an upper bound as explained in the introduction. In our experiments, we have center wavelength,  $\lambda_0= 0.6\thinspace \mu$m, and $\Delta\lambda = 0.4\thinspace \mu$m, which gives $\frac{\Delta\lambda}{\lambda_0}=0.67$, obeying this bound. In prior work, ({\it 24\/}), relative bandwidth close to this bound was achieved at the expense of increased background (and consequent reduction in Strehl ratio).

In conclusion, we showed that it is indeed possible to achieve large aperture (diameter = 100\thinspace mm), relatively low f\# (f/2), flat (thickness = $2.4\thinspace\mu$m) lenses that exhibit a large operating bandwidth (400\thinspace nm to 800\thinspace nm). We experimentally confirmed achromatic focusing with Strehl ratio higher than that of a refractive singlet at its design wavelength. We also demonstrated standalone and hybrid refractive-diffractive telescopes. Due to the extremely low weight, such lenses, especially when scaled to even larger diameters, and when combined with advanced image-processing methods, can enable new space-based telescopes for direct imaging of exoplanets, one of the key challenges emphasized by the recent decadal survey in astronomy and astrophysics ({\it 25\/}).

\paragraph{Acknowledgements} We thank Drs. Stacie Williams, Rohith Chandrasekhar, Augustine Urbas, Philip Hon, Peter Crabtree, and David Monet for fruitful discussions. We also gratefully acknowledge assistance from Heidelberg Instruments (Christian Bach and Dominique Colle), the University of Utah Nanofab (Joseph Spencer and Brian Baker) for fabrication, and from University of Utah Surface Science Lab (Brian van Devener) for metrology assistance. We are grateful to Fernando Vasquez Guevara and Hank Smith for feedback on the manuscript. 

\paragraph{Funding} 
\begin{description}
\item DARPA FA8650-20-C-7020 
\item ONR N00014-19-1-2458
\item NSF 936729
\end{description}

{\bf References and Notes}

\begin{enumerate}
\item D. Apai, T. D. Milster, D. W. Kim, A. Bixel, G. Schneider, B. V. Rackham, R. Liang, and J. Arenberg, {\it Nautilus observatory: a space telescope array based on very large aperture ultralight diffractive optical elements, Proc. SPIE\/} {\bf 11116}, 1111608 (2019).

\item A. V. Nikonorov, M. V. Petrov, S. A. Bibikov, P. Y. Yakimov, V. V. Kutikova, Y. V. Yuzifovich, A. A. Morozov, R. V. Skidanov, and N. L. Kazanskiy, {\it Toward ultralightweight remote sensing with harmonic lenses and convolutional neural networks, IEEE J. Sel. Top. Appl. Earth Observ. Remote Sensing\/} {\bf 11}, 3338–3348 (2018).

\item J. A. Britten, {\it et al, Large-aperture fast multilevel Fresnel zone lenses in glass and ultrathin polymer films for visible and near-infrared applicatios, Appl. Opt.\/} {\bf 53}(11) 2312-2316 (2014). 

\item H. A. MacEwen and J. B. Breckinridge, {\it Large diffractive/refractive apertures for space and airborne telescopes,” Proc. SPIE Sens. Syst. Space Appl.\/} {\bf 8739}, 873904 (2013).

\item S. V. Beryugina, J. R. Kuhn, M. Langlois, G. Moretto, J. Krissansen-Totton, D. Catling, J. L. Grenfell, T. Santi-Temkiv, K. Finster, J. Tarter, F. Marchis, H. Hargitai and D. Apai, {\it The Exo-Life Finder (ELF) telescope: New strategies for direct detection of exoplanet biosignatures and technosignatures, Proc. SPIE\/} v. {\bf 10700}, 1070041 (2018). 

\item M-L. Hsieh, T. D. Ditto, Y-W. Lee, S-H. Lin, H. J. Newberg and S-Y. Lin, {\it Experimental realization of a Fresnel hologram as a super spectral resolution optical element, Sci. Rep.\/} {\bf 11}, 20764 (2021). 

\item A. She, S. Zhang. S. Shian, D. R. Clarke and F. Capasso, {\it Large area metalenses: design, characterization and mass manufacturing, Opt. Exp.\/} {\bf 26}(2) 1573-1585 (2018).

\item F. Presutti and F. Monticone, {\it Focusing on bandwidth: achromatic metalens limits, Optica\/} {\bf 7}(6) 624-631 (2020).

\item T. Stone and N. George, {\it Hybrid diffractive-refractive lenses and achromats,” Appl. Opt.\/} {\bf 27}(14) 2960-2971 (1988).

\item D. W. Sweeney and G. E. Sommargren, {\it  Harmonic diffractive lenses, Appl. Opt.\/} {\bf 34}(14) 2469-2475 (1995).

\item R. Yang, W. Tang and Y. Hao, {\it A broadband zone plate lens from transformation optics, Opt. Exp\/} {\bf 19}(13) 12348-12355 (2011).

\item M. Meem, S. Banerji, A, Majumder, J. C. Garcia, O. Kigner, P. Hon, B. Sensale-Rodriguez and R. Menon, {\it Imaging from the visible to the longwave infrared via an inverse-designed flat lens,” Opt. Exp.\/} {\bf 29}(13) 20715-20723 (2021).

\item Z. Wang, Y. Kim, and T. Milster, {\it High-harmonic diffractive lens color compensation, Appl. Opt.\/} {\bf  60}(19) DD73-D82 (2021).

\item A. Ndao, L. Hsu, J. Ha, J-H. Park, C. Chang-Hasnain and B. Kante, {\it Octave bandwidth photonic fishnet-achromatic-metalens, Nat. Comm. \/} {\bf 11}, 3205 (2020).

\item S. Shreshta, A. C. Overvig, M. Lu, A. Stein and N. Yu, {\it Broadband achromatic dielectric metalenses, Light: Sci \& Appl.\/} {\bf 7}, 85 (2018). 

\item W. T. Chen, A. Y. Zhu, J. Sisler, Z. Bharwani andd F. Capasso, {\it A broadband achromatic polarization-insensitive metalens consisting of anisotropic nanostructures, Nat. Comm.\/} {\bf 10}, 355 (2019).

\item J. W. Goodman, {\it Introduction to Fourier Optics.\/} New York: McGraw-Hill, (1996).

\item S. Banerji and B. Sensale-Rodriguez, {\it A computational design framework for efficient, fabrication error-tolerant planar THz diffractive optical elements, Sci. Rep.\/} {\bf 9}, 5801 (2019).

\item R. Pestourie, C. Pérez-Arancibia, Z. Lin, W. Shin, F. Capasso, and Steven G. Johnson, {\it Inverse design of large-area metasurfaces, Opt. Exp.\/} {\bf 26}(26) 33732-33747 (2018).

\item P. Wang and R. Menon, {\it Computational multi-spectral video imaging, J. Opt. Soc. Am. A\/} {\bf 35}(1), 189-199 (2018).

\item R. Pestourie, C. Pérez-Arancibia, Z. Lin, W. Shin, F. Capasso, and Steven G. Johnson, {\it Inverse design of large-area metasurfaces, Opt. Exp.\/} {\bf 26}(26) 33732-33747 (2018).

\item J. W. Goodman, {\it Statistical Optics.\/} John Wiley \& Sons, 2000.

\item C. Akcay, P. Parrein, and J. P. Rolland, {\it Estimation of longitudinal resolution in optical coherence imaging, Appl. Opt.\/} {\bf 41}(25) 5256-5262 (2002).

\item M. Meem, S. Banerji, A, Majumder, J. C. Garcia, O. Kigner, P. Hon, B. Sensale-Rodriguez and R. Menon, {\it Imaging from the visible to the longwave infrared via an inverse-designed flat lens, Opt. Exp.\/} {\bf 29}(13) 20715-20723 (2021).

\item National Academies of Sciences, Engineering, and Medicine. 2021. {\it Pathways to Discovery in Astronomy and Astrophysics for the 2020s\/}. Washington, DC: The National Academies Press.

\end{enumerate}

\end{document}